\documentclass[aps,prx,twocolumn,nofootinbib]{revtex4-2}

\usepackage{graphicx}% Include figure files
\usepackage{dcolumn}% Align table columns on decimal point
\usepackage{bm}% bold math
\usepackage{textcomp}
\usepackage{color}
\usepackage{amsmath}
\usepackage{amssymb}
\usepackage{xcolor}
\usepackage{hyperref}
\usepackage{easyReview}
\usepackage{mathdots}
\usepackage{float}
\usepackage{lineno}
\usepackage{array}
\usepackage{xcolor,colortbl}
\usepackage[normalem]{ulem}
\definecolor{LightBlue}{rgb}{0.8,0.8,0.8}
\allowdisplaybreaks
\usepackage{amsmath}
\usepackage{lineno}

\begin{document}

\title{Quantum-Inspired Weight-Constrained Neural Network: Reducing Variable Numbers by 100x Compared to Standard Neural Networks}

% Use letters for affiliations, numbers to show equal authorship (if applicable) and to indicate the corresponding author
\author{Shaozhi Li}
\email{lishaozhiphys@mail.neu.edu.cn}
%\affiliation{Department of Physics, Northeastern University, Shenyang 110819, China}
\affiliation{National Center for Transportation Cybersecurity and Resiliency (TracR), Clemson University, SC 29631, USA}
\author{M Sabbir Salek}
\affiliation{National Center for Transportation Cybersecurity and Resiliency (TracR), Clemson University, SC 29631, USA}
\affiliation{Glenn Department of Civil Engineering, Clemson University, Clemson, SC 29634, USA}
\author{Mashrur Chowdhury}
\affiliation{National Center for Transportation Cybersecurity and Resiliency (TracR), Clemson University, SC 29631, USA}
\affiliation{Glenn Department of Civil Engineering, Clemson University, Clemson, SC 29634, USA}
\author{Yao Wang}
\email{yao.wang@emory.edu}
\affiliation{Department of Chemistry, Emory University, Atlanta, GA 30322, USA}

\begin{abstract}
Although quantum machine learning has shown great promise, the practical application of quantum computers remains constrained in the noisy intermediate-scale quantum era. To take advantage of quantum machine learning, we investigate the underlying mathematical principles of these quantum models and find that the quantum neural network with amplitude encoding is equivalent to a weight-constrained neural network. Motived by this discovery, we develop a classical weight-constrained neural network. We find that this approach can reduce the number of variables in a classical neural network by a factor of 135 while preserving its accuracy. In addition, we develop a dropout method to enhance the robustness of quantum machine learning models, which are highly susceptible to adversarial attacks. This technique can also be applied to improve the adversarial robustness of the classical weight-constrained neural network, which is essential for industry applications, such as self-driving vehicles. Our work offers a novel approach to reduce the complexity of large classical neural networks, addressing a critical challenge in machine learning.
\end{abstract}

\maketitle

\section{Introduction}
Large artificial intelligence (AI) models are transforming our education and research, enabling groundbreaking advances in various domains~\cite{Soori2023Artificial,Roberto2023ChatGPT,Arun2023Large,Min2023Recent,Liu2023Summary,Qiu2023Large}. For example, OpenAI's GPT series and Google's BERT have revolutionized our ability to generate human-like texts, recognize complex image patterns, and interpret audio~\cite{devlin2019bert}. Similarly, AlphaFold has simplified the design and discovery of new proteins and drugs~\cite{Jumper2021AlphaFold,Varadi2024AlphaFold}. The remarkable capabilities of these large AI models arise from their vast networks, often containing billions or even trillions of variables~\cite{allenzhu2020learning}. However, the large scale of these AI models induces many significant challenges, including high memory requirements, overfitting, convergence difficulties, and complex hyperparameter tuning~\cite{Erhan2009The,Bachlechner2021Proceedings}. Addressing these challenges requires the development of AI models that could maintain strong learnability while reducing the complexity of the models.

% quantum advantage in decreasing model size
Emerging quantum machine learning techniques offer promising solutions to some challenges posed by large AI models~\cite{Arute2019Quantum,Schuld2022Is,Du2020Expressive,alcazar2022geo,Cerezo2022Challenges,Liu2021A}. The advantage of quantum speed-ups has been demonstrated in various machine learning tasks~\cite{Schuld2022Is,Paparo2014Quantum}, including finding concise functions to fit exponentially large datasets~\cite{Wiebe2012Quantum}, learning a particular data distribution~\cite{Liu2021A}, discovering unknown kernel functions~\cite{Huang2021Power}, and extracting insights from experimental data points using exponentially fewer data compared to classical approaches~\cite{Huang2022Quantum}. These advances can significantly reduce the computational time required to train AI models. In addition to speed, quantum machine learning offers memory advantages by taking advantage of the exponential storage capability of quantum computers. For example, quantum neural networks with amplitude encoding require far fewer variables than their classical counterparts~\cite{salek2024hybrid}. Consequently, the vast number of parameters in large AI models can be effectively reduced by employing quantum or hybrid quantum-classical models.

% application limits
Driven by these advantages, quantum machine learning (QML) has become an active area of research~\cite{RebentrostPRL2014,BiamonteNature2017,ciliberto2018,Beer2020Training,schuld2021supervised, Blanknpj2020, Zoufainpj2019, HavlicekNature2019, Huangnpj, TancaraPRA2023, SlatteryPRA2023}. Although substantial progress has been made in the development of QML algorithms, there are still few real-world applications that leverage real quantum computers~\cite{Benedetti2018Quantum, Perdomo2018opportunities, Rudolph2022Generation, Krunic2022Quantum, moradi2022error}. This gap largely results from the quantum noise issue, which makes results unreliable when many qubits are used~\cite{Terhal2015Quantum,Knill2000Theory}. Addressing this noise challenge is complex, and achieving a large-scale, noise-free quantum computer is expected to take considerable time. In light of these obstacles, the development of quantum-inspired classical algorithms that address the challenges of classical machine learning is a valuable direction to advance machine learning techniques~\cite{Tang2-19Aquantum,Felser2021Quantum,Duong2022Quantum,Ding2022Quantum,Wall2021Tree,li2024quantum,Tiwari2019Towards}.

To translate the advantages of quantum machine learning into classical methods, a deeper understanding of their mathematical foundations is required. Currently, such understanding is limited, and research often focuses on empirically demonstrating the success of QML across various datasets. For instance, quantum neural networks have been show to succeed with fewer parameters in tasks like recovery rate prediction and medical image clarification; however, the reasons for this efficiency remain unclear~\cite{chen2025hybridquantumneuralnetworks,HASSAN2024105560}. Our work aims to clarify the mathematical principles of QNNs and explain the origin of these quantum advantages. We will then leverage this understanding to develop a novel classical neural network, achieving high performance with fewer parameters.

% quantum-inspired classical methods
%The {\it major contribution} of this work is presenting a new method to reduce the number of variables in classical deep-learning neural networks by drawing inspiration from quantum neural networks that employ amplitude encoding. 
We first analyze the mathematical foundations of quantum neural networks with angle encoding and find that using a quantum circuit with angle encoding is mathematically equivalent to constructing a polynomial function, where each term is expressed as a continuous product of the cosine series~\cite{li2024quantum}. In contrast, a quantum circuit with amplitude encoding generates a weight-constrained matrix, whose weight element is again expressed as a summation of the continuous product of the cosine series. The number of trainable variables in this weight-constrained matrix is much smaller than the dimensionality $N$ of the matrix or the input space because the number of quantum gates is typically scaled as $\mathrm{log}_2N$ not $N$. Therefore, a quantum circuit with amplitude encoding requires fewer variables than the classical network, whose variable number is typically scaled as $N$. Motivated by this discovery, we develop a classical weight-constrained neural network, which can reduce the number of variables by 100 times compared to the standard neural network (NN).  We validate this weight-constrained method by examining its performance on four datasets: MNIST, Fashion MNIST (FMNIST), CIFAR, and TRAFFIC datasets. Our results demonstrate that the weight-constrained method maintains the accuracy of the classical neural network while substantially reducing the number of variables. This approach holds promise for scaling down variable numbers in large AI models such as GPT and BERT. Furthermore, we develop a novel method to enhance the adversarial robustness of the quantum neural network and the classical weight-constrained neural network models by randomly dropping parameterized angles during prediction. We demonstrate that such a method can significantly increase robustness against adversarial attacks. 

This paper is organized as follows. Section~\ref{sec:II} presents the mathematical foundations of quantum neural networks with angle and amplitude encodings. Section~\ref{sec:III} develops the theoretical framework of the proposed weight-constrained approach. Section~\ref{sec:IV} demonstrates the effectiveness of applying this method to classical neural networks. Section~\ref{sec:V} compares the performance of weight-constrained neural networks with that of networks employing low-rank matrices. Section~\ref{sec:VI} examines the improved adversarial robustness achieved through our proposed dropout technique. Section~\ref{sec:VII} analyzes the key benefits and broader implications of the weight-constrained methodology. Finally, Section~\ref{sec:VIII} concludes the paper with a summary of findings and potential directions for future research.

\begin{figure}[t]
\centering\includegraphics[width=\columnwidth]{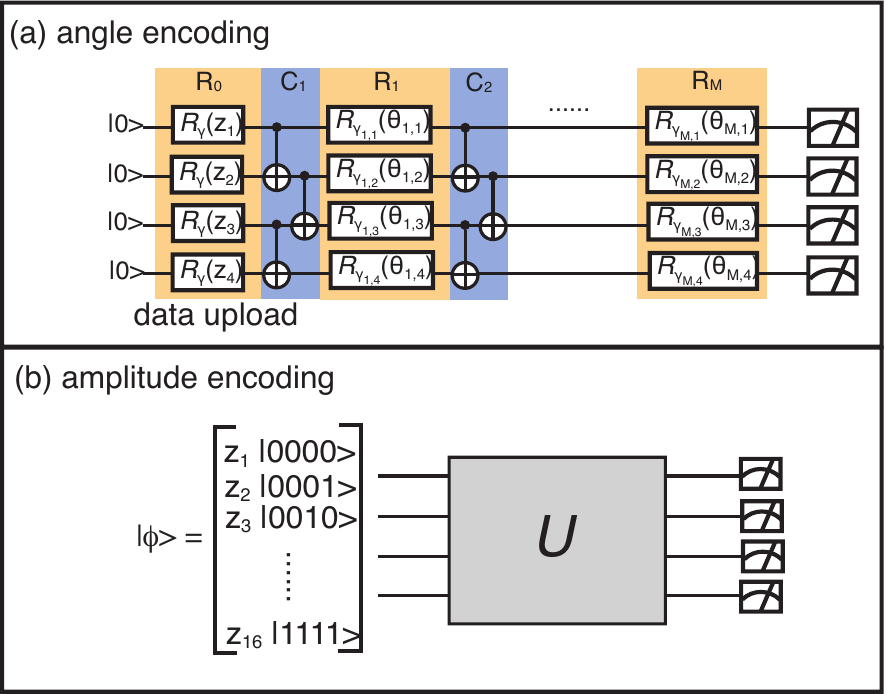}
\caption{Hybrid quantum-classical neural network. 
(a) A generic quantum circuit with angle encoding. (b) A generic quantum circuit with amplitude encoding. 
}
\label{Fig:fig1}
\end{figure}

\section{Quantum neural networks}\label{sec:II}
To develop quantum-inspired classical neural networks, we explore the underlying mathematics of quantum neural networks with angle and amplitude encodings~\cite{Soori2023Artificial}. After building on this mathematical foundation, we develop a quantum-inspired classical neural network, which utilizes 100 times fewer variables than a standard classical neural network.

\subsection*{Quantum neural network with angle encoding} 
We start to analyze the quantum neural network with angle encoding, which adopts the input value as rotation angles in the quantum gates, as shown in Fig.~\ref{Fig:fig1}(a). The nonlinear relationship between the output of a quantum circuit and the input data ${\bf z}$ is constructed by applying different rotation gates and CNOT gates in this circuit. Without loss of generality, we consider a quantum circuit shown in Fig.~\ref{Fig:fig1}(a), which has one data-upload layer $\tilde{R}_0$, $M$ rotation layers, and $M$ CNOT layers. The rotation operator $R_\gamma$ in Fig.~\ref{Fig:fig1}(a) is chosen from the collection of $\{R_x, R_y, R_z\}$.
Then, the final quantum state of this quantum circuit is 
\begin{eqnarray}
|\psi\rangle = \tilde{R}_M \tilde{C}_M \cdots \tilde{R}_1 \tilde{C}_1 \tilde{R}_0 |0000\rangle,
\end{eqnarray}
where $\tilde{R}_i$ denotes a matrix constructed from the tensor product of different single-gate rotation matrices within one layer, and $\tilde{C}_i$ denotes a matrix constructed from the tensor product of two-gate CNOT matrices. Since $\tilde{R}_i$ is composed solely of rotation operators, its matrix elements are either trigonometric functions or zero, whereas the matrix elements of $\tilde{C}_i$ are either 0 or 1. Defining
\begin{eqnarray}
U=\tilde{R}_M \tilde{C}_M \cdots \tilde{R}_1 \tilde{C}_1 \tilde{R}_0,
\end{eqnarray}
then the final state can be expressed as
\begin{eqnarray}
|\psi\rangle = U|0000\rangle=\sum_{i} U_{i,1} |b_i\rangle,
\end{eqnarray}
where $b_i$ denotes the computational basis states (0000, 0001, 0011, ...), and $U_{i,1}$ is the $i$-th element of the first column of $U$, which is either a sum of products of trigonometric functions or zero due to the matrix multiplication. In general, $U_{i,j}$ can be written as
\begin{eqnarray}\label{eq:U}
U_{i,j}=\sum_{s} p_s \prod_{i\in \mathbb{N}_s} \mathrm{cos}(\theta_i/2+q_i) \prod_{n\in \mathbb{M}_s} \mathrm{cos}(z_n/2+q_n),
\end{eqnarray}
where $p_s$ belongs to \{$0, 1,-1,i,-i$\}. The symbol $q_i$ denotes $0$ or $\pi/2$. $\mathbb{N}_s$ and $\mathbb{M}_s$ are collections of integer numbers. 

For a measurement operator $A$ composed of Pauli operators, the circuit output is 
\begin{eqnarray}\label{eq:angle}
O&=&\langle \psi | A | \psi \rangle =\sum_{i,j}U^{*}_{i,1}A_{i,j}U_{j,1}=\sum_{s}f_s({\bf \theta},{\bf z}),\\
f_s({\bf \theta},{\bf z}) &=& c_s \prod_{i\in \mathbb{N}_{1,s}} \mathrm{cos}(\theta_i/2+q_i) \prod_{j\in \mathbb{N}_{2,s}} \mathrm{cos}(\theta_j/2+q_j)  \times \nonumber\\
&&\prod_{n\in \mathbb{M}_{1,s}} \mathrm{cos}(z_n/2+q_n) \prod_{m\in \mathbb{M}_{2,s}} \mathrm{cos}(z_m/2+q_m),\nonumber\\
\end{eqnarray}
where $c_s$ belongs to \{$1,-1$\}. It is straightforward to find that the output $O$ of a quantum circuit is a polynomial function. Equation~(\ref{eq:angle}) shows that the quantum neural network with angle encoding generates a complicated activation function, which could select features better than the classical activation functions, such as ReLU and Tanh. We refer the reader to our previous work~\cite{li2024quantum} for a deeper understanding of this quantum activation function.

\subsection*{Quantum neural network with amplitude encoding is a weight-constrained neural network}
Unlike angle encoding, which transforms real-world data into rotation angles, amplitude encoding transforms real-world data into coefficients of pure quantum states. For the amplitude encoding, the initial state of a quantum circuit $|\phi\rangle$ is set as
\begin{eqnarray}
|\phi\rangle_i = \frac{z_i}{|{\bf z}|} |b_i\rangle,
\end{eqnarray}
where $|b_i\rangle$ denotes a pure quantum state, i.e., $|0000\rangle$ and $|0001\rangle$. $|{\bf z}|$ is the norm value of the input vector ${\bf z}$. This initial quantum state can be transformed to other states by applying rotation and CNOT gates. For simplicity, we can use an operator $U$ to represent these operations as shown in Fig.~\ref{Fig:fig1}(b). Then, the final quantum state is written as
\begin{eqnarray}
|\psi\rangle=U|\phi\rangle.
\end{eqnarray}
If we set the measurement operator as $A$, the output $O$ of this quantum circuit is
\begin{eqnarray}\label{eq:amplitude}
O=\langle \phi | U^{\dagger} A U |\phi \rangle=\langle \phi | W|\phi \rangle,
\end{eqnarray}
where $W$ is a unitary matrix. It is easy to rewrite Eq.~(\ref{eq:amplitude}) as 
\begin{eqnarray}\label{eq:amplitude_expand}
O=\frac{1}{|{\bf z}|^2}\sum_{i,j} z_i W_{ij} z_{j}.
\end{eqnarray}
Equation~(\ref{eq:amplitude_expand}) shows that the quantum circuit generates a correlation matrix between any two elements in the input data. For example, $W_{1,2}$ represents the correlation between the first and second input features, and $W_{1,100}$ denotes the correlation between the first and the one hundredth input features. If all elements in $W$ are nonzero, the construction of features incorporates long-range and short-range correlations\cite{Huang2021Power}.

The weight matrix element $W_{ij}$ is mathematically equivalent to the matrix $U$ defined in Eq.~(\ref{eq:U}). This weight element is given by
\begin{eqnarray}\label{eq:wij}
W_{i,j}=\sum_s p_s \prod_{k \in \mathbb{N}_s} \mathrm{cos}(\theta_k/2+q_k), 
\end{eqnarray}
where $\theta_k$ is a rotation angle, $p_s \in \{0, 1,-1,i,-i \}$. $q_k \in \{ 0, \pi/2 \}$, and $\mathbb{N}_s$ is a set of integers.

Equation~(\ref{eq:wij}) shows that each weight element is a polynomial function, where each term is a continuous product of the cosine series. Therefore, {\it the amplitude encoded quantum neural network is a weight-constrained neural network.} In a typical quantum circuit, the number of rotation angles is significantly smaller than the dimensionality of the Fock space it encodes. Consequently, an amplitude-encoded quantum neural network operates with fewer trainable variables than a standard neural network of equivalent representational capacity.

\section{Quantum-inspired classical network: weight-constrained neural network}\label{sec:III}
Motivated by the discovery that the amplitude encoded quantum neural network is a weight-constrained neural network, we develop a classical counterpart of the weight-constrained neural network, in which the weight variables are constructed as follows. To generate $K$ different weights, we first define $N$ different variables, denoted as $\theta$. Then a combination process is performed, where $r$ variables are chosen from these $N$ variables, resulting in $C(N,r)$ possible combinations. Labeling the $k$-th combination as ${\bf \theta}^{(k)}$, the $k$-th weight $w_k$ is defined as
\begin{eqnarray}\label{eq:consw}
w_k=\prod_{i\in \mathrm{even}, i\le r} \mathrm{cos}(\theta_i^{(k)}) \prod_{j\in \mathrm{odd}, j\le r} \mathrm{sin}(\theta_j^{(k)}).
\end{eqnarray}
This combination strategy enables constructing thousands of weights using only a few dozen variables. We note that the weight correlation at the first moment is zero because
\begin{eqnarray}
\int_{-\pi}^{\pi} d{\bf \theta} w_k w_{k^\prime} = 2^{N-r}\pi^N\delta_{k,k^\prime}.
\end{eqnarray}
In addition, these weights are always correlated in higher moments, that is
\begin{eqnarray}
\int_{-\pi}^{\pi} d\theta w_k^s w_{k^\prime}^s \ne 0,
\end{eqnarray}
when $s>1$. 

A critical question when adopting the weight-constrained method is how it impacts the performance of a neural network. To answer this question, we examine two distinct two-layer neural networks. In the first classical network, 100 input features, $x$, are connected to an output neuron, $y$, as follows:
\begin{eqnarray}
y&=&f(z),\\
z&=&\sum_{i} w_ix_i\label{eq:z}
\end{eqnarray}
where $w_i$ represents the weight, sampled from the uniform distribution $\mathcal{U}(-1,1)$. Here, $f(z)$ denotes a standard classical activation function, such as Tanh, Sigmoid, or ReLU. The second classical neural network has the same architecture as the first but employs the weight-constrained method to determine the weights. Since the variance of the constrained weights, as defined in Eq.~(\ref{eq:consw}), scales as $2^{-r/2}$, we adjust them to remove the impact of the variance. We redefine the weights as: 
\begin{eqnarray}\label{eq:consw2}
w^\prime_k=\sqrt{\frac{2^r}{3}}\prod_{i\in \mathrm{even}, i\le r} \mathrm{cos}(\theta_i^{(k)}) \prod_{j\in \mathrm{odd}, j\le r} \mathrm{sin}(\theta_j^{(k)}).
\end{eqnarray}
Then the variance of $w^\prime_k$ is $\sqrt{1/3}$, matching the variance of the weights $w_i$ drawn from $\mathcal{U}(-1,1)$.

\begin{figure}[t]
\centering\includegraphics[width=\columnwidth]{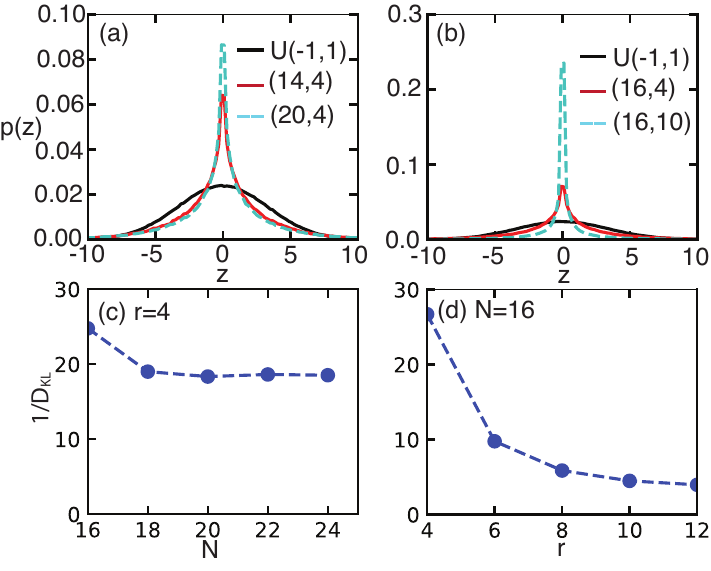}
\caption{ Panel (a) and panel (b) plot the distribution $p(z)$ of $z$, defined as $z=\sum_i w_i x_i$, where $w$ denotes the weight, and $x$ denotes the input features. (c) The inverse of Kullback-Leibler divergence $D_\text{KL}^{-1}$ of the distribution $p(z)$ as a function of $N$ at $r=4$, where $N$ denotes the number of variables. (d) $D_\text{KL}^{-1}$ of the distribution $p(z)$ as a function of $r$ at $N=16$.}
\label{Fig:fig2}
\end{figure}

\begin{figure*}
\centering\includegraphics[width=\textwidth]{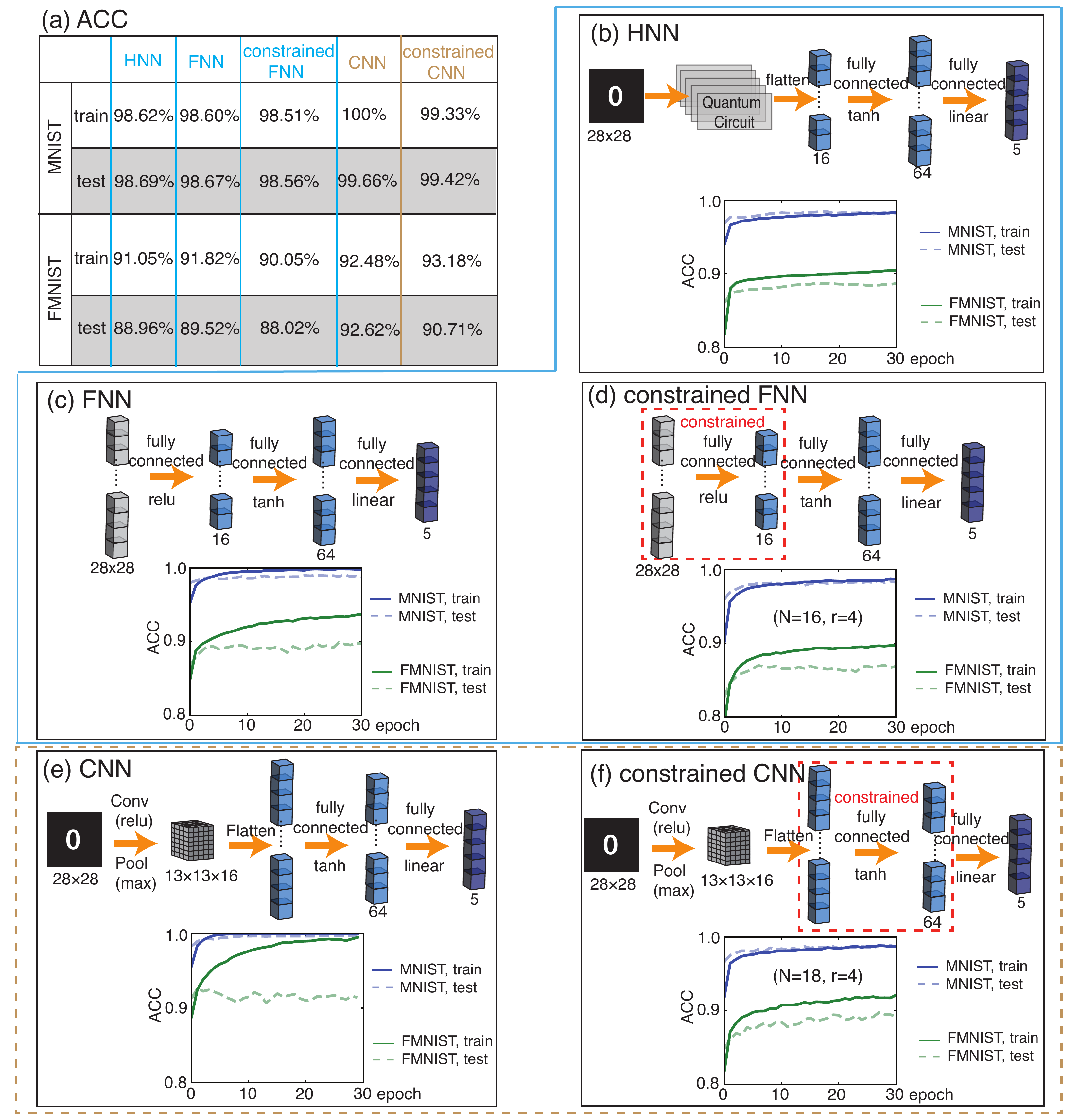}
\caption{Training results of different neural networks. (a) A summary of the accuracy (ACC) of these five models for MNIST and FMNIST datasets. (b) The architecture of the hybrid quantum-classical neuron network (HNN) and the training results on the MNIST and Fashion MNIST (FMNIST) datasets. (c) The fully connected neuron network (FNN) architecture and the training results on the MNIST and FMNIST datasets. (d) The weight-constrained FNN architecture and the training results on the MNIST and FMNIST datasets. (e) The convolutional neural network (CNN) architecture and the training results on the MNIST and FMNIST datasets. (f) The architecture of the weight-constrained CNN and the training results on the MNIST and FMNIST datasets. 
}
\label{Fig:fig3}
\end{figure*}

Figures~\ref{Fig:fig2}(a) and~\ref{Fig:fig2}(b) display the distribution $p(z)$ of $z$ for both classical neural networks. The input features $x$ are sampled from a uniform distribution $\mathcal{U}(-1,1)$. The black curve represents the distribution for the classical neural network with weights sampled from $w\in \mathcal{U}(-1,1)$. The other curves show results for the weight-constrained network across different pairs of $(N, r)$ used in constructing the combination sets. As shown in Figs.~\ref{Fig:fig2}(a) and~\ref{Fig:fig2}(b), both networks exhibit dome-shaped distributions centered at $z=0$. Including a uniformly distributed bias term does not affect the central position of these distributions. 

Although the mean and variance of the distributions for the weights from the uniform distribution and the weight-constrained method are the same, the weight-constrained network displays a narrower distribution width and a more pronounced peak compared to the network using uniformly distributed weights. Moreover, we observe that the weight distribution of the weight-constrained network shows a strong dependence on $r$ but a weak dependence of $N$. For instance, at $r=4$, increasing the value of $N$ from 14 to 20 only slightly raises $p(z=0)$ from 0.06 to 0.08. In contrast, at $N=16$, increasing $r$ from 4 to 10 causes a significant enhancement of $p(z=0)$. This result implies that the performance of the weight-constrained network has a strong dependence on $r$ but a weak dependence of $N$.

To quantify the performance of a network, we evaluate its expressibility, which characterizes the ability of a network to represent or predict arbitrary real value. In the ideal case, the variable $z$ can take any real value within the range $(-\infty, \infty)$, and its probability distribution $p(z)$ is uniform. Such a case corresponds to the strongest expressibility. However, when constraints are imposed on both the weights $w$ and inputs $x$, the resulting distribution $p(z)$ becomes dependent on $z$, thereby reducing the expressibility. To quantify this deviation, we employ the Kullback-Leibler (KL) divergence, $D_{\mathrm{KL}}$, which measures the difference between $p(z)$ of the network and $q(z)$ of the ideal case:
\begin{eqnarray} 
D_\text{KL} = \int_{-\infty}^{\infty} dz p(z) \log\left[ \frac{p(z)}{q(z)}\right].
\end{eqnarray}
In practical computations, we should restrict the integration domain to $(-R,R)$ with a large cutoff $R$, for which $q(z)=1/(2R)$. Under this approximation, the KL divergence becomes 
\begin{eqnarray}
D_\text{KL} = \int_{-R}^R dz p(z) \log p(z) + \log (2R),
\end{eqnarray}
when $p(z)$ closely resembles the uniform distribution of the ideal case, $D_\text{KL} \approx 0$; conversely, when $p(z)$ collapses to a delta function $\delta(z)$, we obtain $D_\text{KL}=\log (2R)$. Therefore, $D_\text{KL}$ serves as an inverse measure of expressibility — the smaller the KL divergence, the greater the expressibility of a network.

In our calculations, we use the output distribution $p(z)$ of a network with weights sampled from $U(-1,1)$  as the reference distribution $q(z)$. 
This choice is justified because this unrestricted network possesses higher expressibility than the weight-constrained network (see Appendix F for details). Figures~\ref{Fig:fig2}(c) and~\ref{Fig:fig2}(d) plot the inverse KL divergence, $1/D_\text{KL}$ (blue circles), as a function of $N$ (at $r=4$) and $r$ (at $N=16$), respectively. The results indicate that $1/D_\text{KL}$ decreases slightly with increasing $N$ for $N<18$ and becomes independent of $N$ for $N>18$. In contrast, $1/D_\text{KL}$ decreases rapidly as $r$ increases. This behavior suggests that the expressibility of the weight-constrained network depends weakly on $N$, while it increases significantly as $r$ decreases. 
%Furthermore, we observe that $1/D_\text{KL}$ of uniformly distributed weights is larger than that of the weight-constrained network, indicating that the standard network exhibits higher expressibility than the weight-constrained network.

\section{Practicality of the weight-constrained method }\label{sec:IV}
\begin{figure*}
\centering\includegraphics[width=\textwidth]{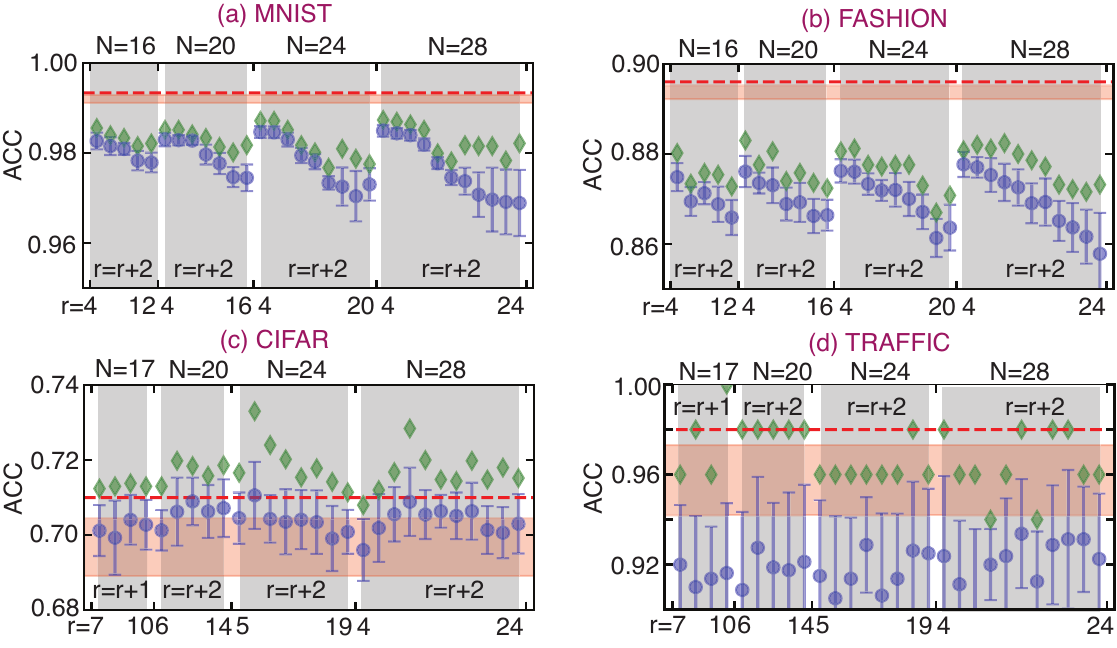}
\caption{The accuracy (ACC) of the weight-constrained fully connected neural network (FNN) and the weight-constrained convolutional neural network (CNN) for different values of $N$ and $r$. Here, $N$ and $r$ represent variables in the combination formula $C(N,r)$. Panel (a) and panel (b) plot the results of the MNIST and the FMNIST datasets. Panel (c) and panel (d) plot the results of the CIFAR and the TRAFFIC datasets. The red dashed line and shade region represent the optimal value and the $\pm1$ standard deviation interval of the standard network, respectively. 
}
\label{Fig:fig4}
\end{figure*}

Having discussed the expressibility of classical neural networks employing the weight-constrained method, we now focus on its practical applications in classical fully connected and convolutional neural networks.

We begin by comparing the performance of the weight-constrained neural network with that of a standard classical fully connected neural network (FNN) and a hybrid neural network (HNN). The architectures of these three networks are depicted in Figs.~\ref{Fig:fig3}(b)-~\ref{Fig:fig3}(d), respectively. All three networks consist of four layers, with 16 neurons in the first hidden layer and 64 neurons in the second hidden layer. Details on the quantum circuit design are provided in Appendix A. For the weight-constrained FNN, the weights in the first hidden layer are determined using Eq.(\ref{eq:consw2}), which leads to the same variance for different $r$ values. In addition to FNN, we evaluate the performance of the weight-constrained convolutional neural network (CNN) and its standard counterpart. Figure~\ref{Fig:fig3}(e) illustrates a standard CNN architecture comprising one convolutional layer, one pooling layer, and two fully connected layers. The weight-constrained CNN, shown in Fig.~\ref{Fig:fig3}(f), shares the same architecture as the standard CNN, but the weight-constrained method is applied to the first fully connected layer. The weights for the standard layer are initialized using the He method, and the variables for the weight-constrained method are initialized from a uniform distribution $\mathcal{U}(-\pi,\pi)$.

The table in Fig.~\ref{Fig:fig3}(a) summarizes the optimal results of the HNN, FNN, constrained FNN, CNN, and constrained CNN for the MNIST and FMNIST datasets. These results are determined based on the minimal loss function for the test datasets in one training simulation. In the HNN, each quantum circuit comprises 38 variables. In the weight-constrained FNN, the combination sets are constructed by selecting 4 elements from a pool of  16 variables ($N=16$ and $r=4$. This combination set is chosen randomly). With this parameter setting, the number of variables in the first layer of the HNN and the constrained FNN is reduced by factors of approximately 19 and 49, respectively, compared to the standard FNN. For the MNIST test dataset, the HNN and the FNN achieved similar accuracies (ACC), 98.7\%, about 0.1\% larger than the ACC of the weight-constrained FNN. For the FMNIST test dataset, the HNN and the FNN performed comparably, achieving an ACC of 89\%, approximately 1\% larger than that of the weight-constrained FNN. These results indicate that introducing weight constraints does not significantly degrade the performance of neural networks. The slightly better performance of the HNN compared to the weight-constrained FNN can be attributed to the more effective weight construction provided by the quantum circuit.

Figures~\ref{Fig:fig3}(e) and~\ref{Fig:fig3}(f) plot training results of the CNN and the weight-constrained CNN on these two datasets. Here, the weight-constrained method is implemented by selecting 4 elements from a pool of 18 variables, reducing the number of variables by a factor of 150 compared to the standard CNN. Despite this substantial reduction in variables, the table in Fig.~\ref{Fig:fig3}(a) shows that the weight-constrained CNN achieves an ACC comparable to the standard CNN on the MNIST test dataset. For the FMNIST test dataset, the ACC of the weight-constrained CNN is only 1.91\% smaller than that of the standard CNN. Notably, we observe that the loss function of the weight-constrained CNN for the test dataset decreases monotonically, whereas that of the standard CNN first decreases and then increases as the number of epochs grows (see Appendix~\ref{sec:cnnloss} for details). This behavior indicates that the weight-constrained method effectively mitigates overfitting, as the reduced expressibility of the constrained network limits its tendency to learn noise from the training data.

To further investigate the impact of the combination method in the weight-constrained strategy, we perform simulations using different values of $N$ and $r$. Figure~\ref{Fig:fig4} presents the ACC for the MNIST, FMNIST, CIFAR, and TRAFFIC test datasets. In these simulations, the weight-constrained FNN is applied to the MNIST and FMNIST datasets, whereas the weight-constrained CNN is used for the CIFAR and TRAFFIC datasets.  For each $(N, r)$ pair, the green diamond and blue circle (with error bars) represent the optimal and mean ACC from 16 independent simulations, respectively. The red dashed line and shade region represent the optimal value and the $\pm1$ standard deviation interval of the standard network, respectively. For the weight-constrained FNN, the mean ACC generally decreases as $r$ increases, which is attributed to that the expressibility is suppressed by $r$. In contrast, the mean ACC shows no significant dependence on $N$, consistent with our observation that the expressibility depends only weakly on $N$. 
We note that the impact of expressibility is also dependent on the activation function. For the ReLU activation function, $\mathrm{relu}(z)$ is nonzero as $z\rightarrow \infty$, whereas $\mathrm{tanh}(z)$ becomes 0 as $z \gg 1$. Therefore, varying $r$ has a less pronounced impact when the Tanh activation function is applied to the weight-constrained layer (see Appendix~\ref{sec:mnist} for details).  For the weight-constrained CNN, the mean ACC shows no clear relationship with $r$. This is because the neuron value $x$ in the weight-constrained layer can be adjusted by the preceding layers, thereby mitigating the expressivity limitations imposed by the weight-constrained approach (see Appendix~\ref{sec:mnist} for details).

The performance of the weight-constrained neural network, compared to the standard classical neural network, varies depending on the architecture of the neural network. We observe that the optimal ACC of the weight-constrained FNN (green diamonds) is slightly smaller than that of the standard ACC (red dashed line). In contrast, the optimal ACC of the weight-constrained CNN is comparable to or even larger. This result implies that the weight-constrained approach delivers superior performance in CNNs compared to FNNs.

\begin{figure}[t]
\centering\includegraphics[width=\columnwidth]{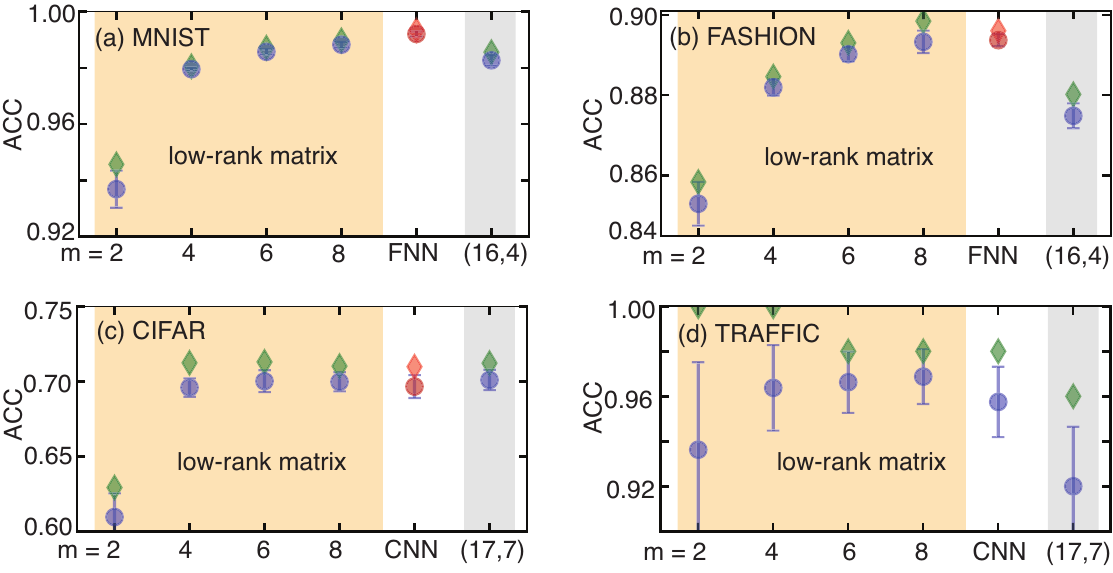}
\caption{Test accuracy (ACC) of neural networks with different weight parameterizations. (a) MNIST and (b) FMNIST results for fully-connected neural networks (FNNs). (c) CIFAR and (d) TRAFFIC dataset results for convolutional neural networks (CNNs). For each panel, performance is shown for a standard network, our proposed weight-constrained method, and a low-rank method with varying $m$.  Markers indicate the mean ACC over 16 simulations (circles) and the optimal result (diamonds).}
\label{Fig:fig5}
\end{figure}

\section{Comparison: weight-constrained vs. low-rank neural networks}\label{sec:V}
Compared to standard neural network, the proposed weight-constrained method has fewer parameters by restricting weight variables. In machine learning, other techniques have been proposed to reduce variables, such as the low-rank matrix~\cite{sun2018igcv} and the weight sharing technique~\cite{lin2023understanding}. Here, we compare the performance of the weight-constrained method with low-rank neural networks. 

A low-rank weight matrix is constructed as $W=A\times B$, where $A$ has dimensions $s\times m$ and $B$ has dimensions $m\times n$. 
The variable number of this low-rank weight matrix is $sm+mn$, which is smaller than the size of $W$ when $m\ll s$ and $m\ll n$; therefore, the low-rank weight matrix method reduces the variable number. To ensue a fair comparison with the weight-constrained neural network, in our study, the low-rank weight matrix is applied in the first hidden layer for the FNN and the second-to-last layer for the CNN.

Figure~\ref{Fig:fig5} shows the ACC for various neural networks across different test datasets. Results show the mean (blue circle) and standard deviation (green diamond) from 16 independent simulations. The red symbol denotes the result of the standard neural network. For the MNIST and FMNIST datasets, we compare a standard FNN against its low-rank matrix and weight-constrained variants. Similarly, for the CIFAR and TRAFFIC datasets, we compare a standard CNN against its low-rank and weight-constrained counterparts. The weight-constrained FNN with $N=16$ and $r=4$ contains 256 parameters in the constrained layer, whereas the low-rank CNN with $N=17$ and $r=7$ includes 1088 parameters. In contrast, the low-rank FNNs with $m=2, 4, 6$, and $8$ contain 1600–6400 parameters in the low-rank matrix, while the corresponding low-rank CNNs contain 21728–86912 parameters. The results show that the mean ACC of the low-rank models increases with $m$, eventually approaching that of the standard NN. Despite using substantially fewer parameters, the weight-constrained networks achieve mean accuracies comparable to those of the low-rank models with $m=4$ on all datasets except TRAFFIC. The deviation observed for the TRAFFIC dataset may result from its relatively small training dataset. Overall, these results demonstrate that the proposed weight-constrained method is more parameter-efficient while maintaining competitive performance relative to the low-rank approach.

\begin{figure*}[ht]
\centering\includegraphics[width=\textwidth]{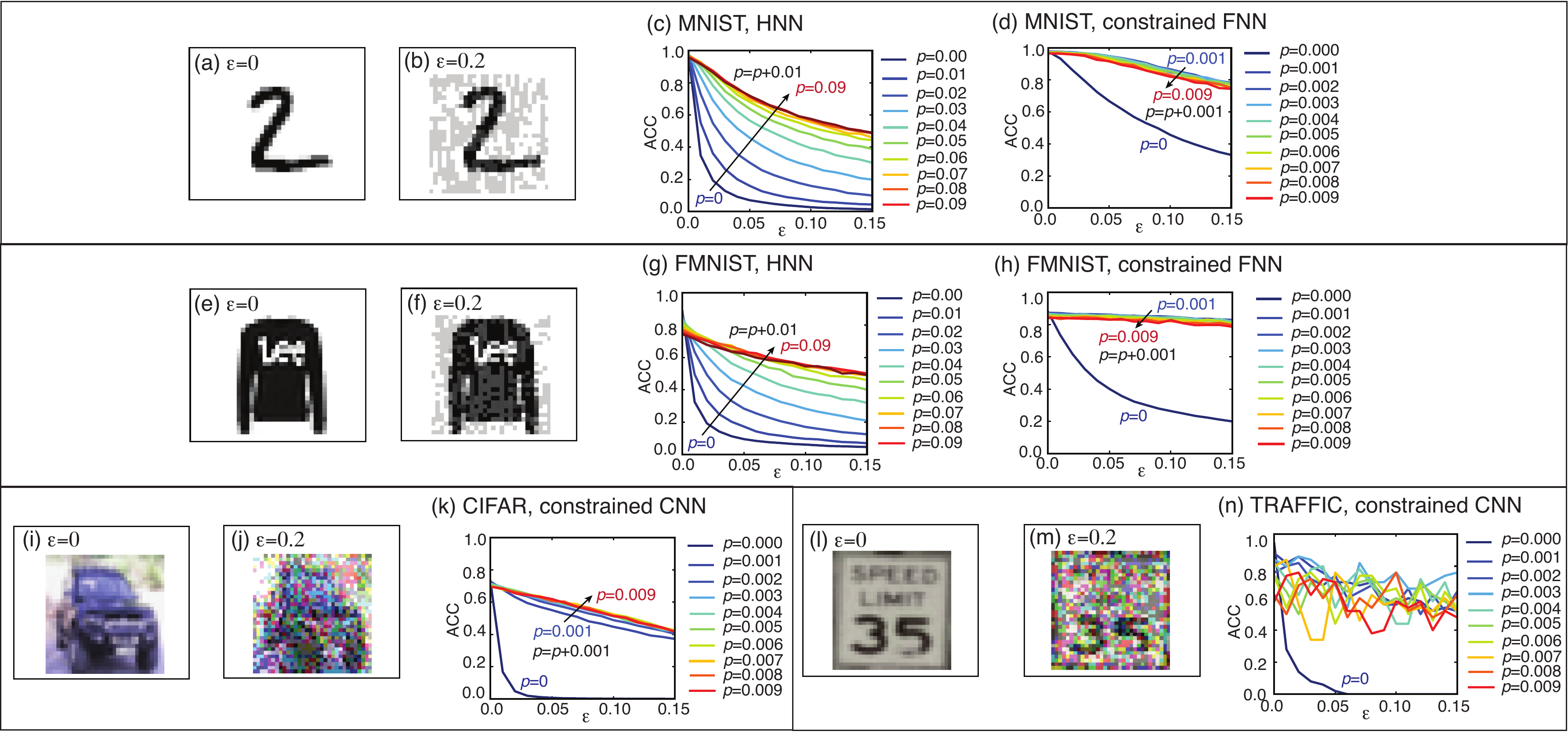}
\caption{The accuracy (ACC) under adversarial attack. The top panel shows the MNIST image and the image under an adversarial attack with an attack intensity $\epsilon=0.2$. Panels (c) and (d) plot the ACC of the hybrid quantum-classical neural network (HNN) and the weight-constrained fully connected neural network (FNN) with a dropout policy. Here, $p$ denotes the dropout probability. The middle panel shows the Fashion MNIST (FMNIST) image and the image under an adversarial attack with an attack intensity $\epsilon=0.2$. Panels (g) and (h) plot the ACC of the HNN and the weight-constrained FNN with a dropout policy. The bottom left panel shows the CIFAR image and the image under an adversarial attack with an attack intensity $\epsilon=0.2$. Panel (k) plots the ACC of the weight-constrained convolutional neural network (CNN) with a dropout policy. The bottom right panel shows the traffic image and the image under an adversarial attack with an attack intensity $\epsilon=0.2$. Panel (n) plots the ACC of the weight-constrained CNN with a dropout policy.
}
\label{Fig:fig6}
\end{figure*}

\section{Dropout enhanced adversarial robustness}\label{sec:VI}
Adversarial machine learning is an emerging frontier that studies the vulnerability of machine learning systems and develops defense strategies against adversarial attacks. It was demonstrated that quantum machine learning methods for classification tasks are highly vulnerable to adversarial attacks as a result of the concentration of measure phenomenon, which refers to that the distance from a (Haar) randomly sampled point in the Hilbert space to the nearest adversarial example vanishes as $\mathcal{O}(2^{-n})$, where $n$ is the number of qubits~\cite{Liu2020PRA}. Moreover, it is found that universal adversarial samples can deceive all classifiers with only a perturbation of strength $\mathcal{O}(\mathrm{log}(k)2^{-n})$, where $k$ is the number of independent classifiers~\cite{weiyuan2022universal}. Recently, several approaches have been proposed to improve the robustness of quantum machine learning models, including applying quantum noise~\cite{du2021quantum} and adversarial training~\cite{lu2020quantum}. In this work, we propose a new approach to improve adversarial robustness.

In this study, we use the fast gradient sign method (FGSM) to generate adversarial samples~\cite{goodfellow2015explaining}. For the FGSM method, each pixel is modified by $\epsilon \mathrm{sgn}(\nabla_x L)$, where $L$ is the loss function and $\epsilon$ represents the strength of the attack. Therefore, a positive value of $\epsilon$ increases the loss function, reducing accuracy. To visualize the adversarial attack, we plot original images and their corresponding adversarially perturbed versions in Figs.~\ref{Fig:fig6}(a),(b),(e),(f),(i),(j),(l), and (m).

To enhance the robustness of the quantum machine learning model, we propose a randomized dropout strategy for $R_z$ gates within the quantum circuit of a trained model. In this approach, dropout is applied after training, and attackers only interact with the model after dropout has been incorporated. Figures~\ref{Fig:fig6}(c) and~\ref{Fig:fig6}(g) plot the ACC of the HNN model under adversarial attack for the MNIST and FMNIST test datasets, where $p$ denotes the probability of dropping out. Without the dropout policy ($p=0$), the ACC for both datasets collapses to near zero at $\epsilon=0.05$, revealing the inherent vulnerability of quantum models to attacks. In contrast, dropout maintains robust performance. For example, with $p=0.08$, the ACC for the MNIST dataset declines from 0.968 to 0.567, and for the FMNIST dataset, it drops from 0.774 to 0.555 when $\epsilon$ increases from 0 to 0.1. It is also observed that the ACC for unattacked samples is suppressed by the dropout probability. For example, increasing the probability of dropping out from 0 to 0.08 reduces the ACC by approximately 0.02 for the MNIST dataset and by 0.112 for the FMNIST dataset. These results highlight that while the dropout method effectively enhances the adversarial robustness of quantum neural networks, it is crucial to carefully select the dropout probability to strike a balance between maintaining ACC for unattacked samples and improving robustness under attack.

The dropout method can be extended to classical weight-constrained neural networks to enhance their adversarial robustness. Dropping an $R_z$ gate in a quantum circuit analogously corresponds to changing elements of the weight matrix $W$ shown in Eq.~(\ref{eq:amplitude_expand}). In a weight-constrained neural network, the weight is constructed from the continuous product of the trigonometric function. Therefore, we implement dropout by randomly selecting rotation angles and setting their corresponding trigonometric terms to 1 (equivalent to removing their contributions). Figure~\ref{Fig:fig6}(d) and Figure~\ref{Fig:fig6}(h) illustrate the ACC of the weight-constrained FNN on the MNIST and FMNIST test datasets under adversarial attack, respectively. It is found that a small nonzero dropout probability ($p=0.001$) can significantly improve adversarial robustness and make the change of the ACC under attacks tiny. For example, when $p=0.001$ is adopted, the ACC for the FMNIST dataset is only decreased by 0.04 as the attack strength $\epsilon$ increases from 0 to 0.15. Compared to the hybrid quantum-classical model, the weight-constrained FNN requires a smaller value of $p$ to achieve enhanced adversarial robustness. This ensures that the ACC for unattacked samples remains almost unaffected when such a small $p$ is adopted. Furthermore, Figs.~\ref{Fig:fig6}(k) and~\ref{Fig:fig6}(n) plot the ACC of the weight-constrained CNN, which is trained on the CIFAR and TRAFFIC datasets, respectively. Again, we observe that the adversarial robustness is enhanced by the dropout method. The oscillatory behavior observed in Fig.~\ref{Fig:fig6}(n) is attributed to the small sample size of the TRAFFIC test dataset (50 samples), which lacks sufficient statistical robustness. 

We note that the success of our approach on the TRAFFIC dataset is critical to ensuring the safety of self-driving vehicles~\cite{Chowdhury2020Attacks,Deng2020An}. In these vehicles, CNN-based regression models are widely used for object recognition. However, CNNs have been shown to be highly susceptible to adversarial samples~\cite{Ian2015Explaining}. To address this issue, numerous studies have focused on improving the adversarial robustness of CNNs~\cite{Papernot2016Distillation,Wong2018Provable}. Our work introduces a novel method to improve CNN robustness, thus improving the safety and reliability of self-driving vehicles.

\section{Overview of the benefits of the weight-constrained neural network}\label{sec:VII}
In machine learning, regularization and early stopping are commonly employed to mitigate overfitting. Our developed weight-constrained neural network introduces an alternative approach to addressing this issue. By limiting its expressiblity, the weight-constrained neural network effectively filters out noise in training data while capturing essential signals, thereby suppressing overfitting (see details in Appendix~\ref{sec:cnnloss}).

Beyond suppressing overfitting, the weight-constrained method significantly reduces memory and storage requirements. In standard neural networks, all weights must be stored in memory, causing memory usage to scale linearly with the number of weights. In addition, large AI models require substantial storage to save trained parameters. The weight-constrained method can reduce these costs. For example, memory usage can be optimized by temporarily constructing and releasing the weight matrix after use. In addition, the memory cost for storing gradients of weights can also be reduced by adopting the weight-constrained method. This reduction is particularly critical for large-scale AI models. For example, a model with one billion parameters requires 8 GB of memory for weights and corresponding gradients with float-32 precision. By applying the weight-constrained method, the number of variables can be reduced by a factor of 100, reducing the memory requirement to just 81 MB.

To effectively apply QML to industry challenges, ensuring the security of QML models is paramount. Although standard classical techniques, such as adversarial training~\cite{lu2020quantum}, can enhance the robustness of QML models, our work introduces an alternative approach that takes advantage of the intrinsic properties of quantum circuits. Additionally, our developed dropout method, based on the weight-constrained method, enriches techniques to improve the robustness of classical machine learning models.

\section{Conclusion}\label{sec:VIII}
In this work, we investiagte the fundamental mathematical principles of quantum neural networks with angle and amplitude encodings. We find that quantum neural network with amplitude encoding is equivalent to a weight-constrained neural network. Building on this insight, we develop a classical weight-constrained neural network that simultaneously addresses overfitting and reduces memory requirements. Furthermore, we develop a dropout approach to enhance the adversarial robustness of both quantum neural networks and classical weight-constrained neural networks. These advancements hold significant promise for practical machine learning applications in industrial settings.

While our study demonstrates the effectiveness of the weight-constrained approach for conventional neural architectures, we have not evaluated its performance on transformer models. Recent work by Cherrat at el.~\cite{Cherrat2024Quantum} has proposed a quantum transformer variant that employs amplitude encoding for attention coefficient computation, demonstrating quantum advantages in MedMNIST classification tasks. This suggests an important direction for future research: investigating how weight constraints might be integrated into transformer architectures. Moreover, we observe that the quantum neural network with amplitude encoding, which corresponds to a quantum weight-constrained neural network, outperforms our proposed classical weight-constrained neural network. This finding encourages further investigation into alternative and potentially more effective weight-constrained methods.

As noted in Ref.~\cite{Wplpert1996The}, there exists an inherent trade-off in algorithm design - better performance on one task often comes at the expense of another. Our proposed weight-constrained method effectively reduces memory usage; however, it increases computational time due to the dynamic memory allocation required for constructing weights. Striking a balance between memory consumption and computational efficiency requires determining the optimal number of weights to be constructed using this method. However, we note that the need for dynamically constructing weight matrices is not required on quantum computers; therefore, quantum machine learning has advantages in both reducing memory consumption and improving computational efficiency. 

\section*{Acknowledgement}
This work is supported by the National Center for Transportation Cybersecurity and Resiliency (TraCR) (a U.S. Department of Transportation National University Transportation Center) headquartered at Clemson University, Clemson, South Carolina, USA. Y.W. acknowledges support from the U.S. Department of Energy, Office of Science, Basic Energy Sciences, under Early Career Award No.~DE-SC0024524. 

{\it Data Availability.} The data that support the findings of this article are not publicly available upon publication because it is not technically feasible and/or the cost of preparing, depositing, and hosting the data would be prohibitive within the terms of this research project. The data are available from the authors upon reasonable request.

\appendix
\section{Quantum Circuit}
\begin{figure}[h]
\centering\includegraphics[width=0.99\columnwidth]{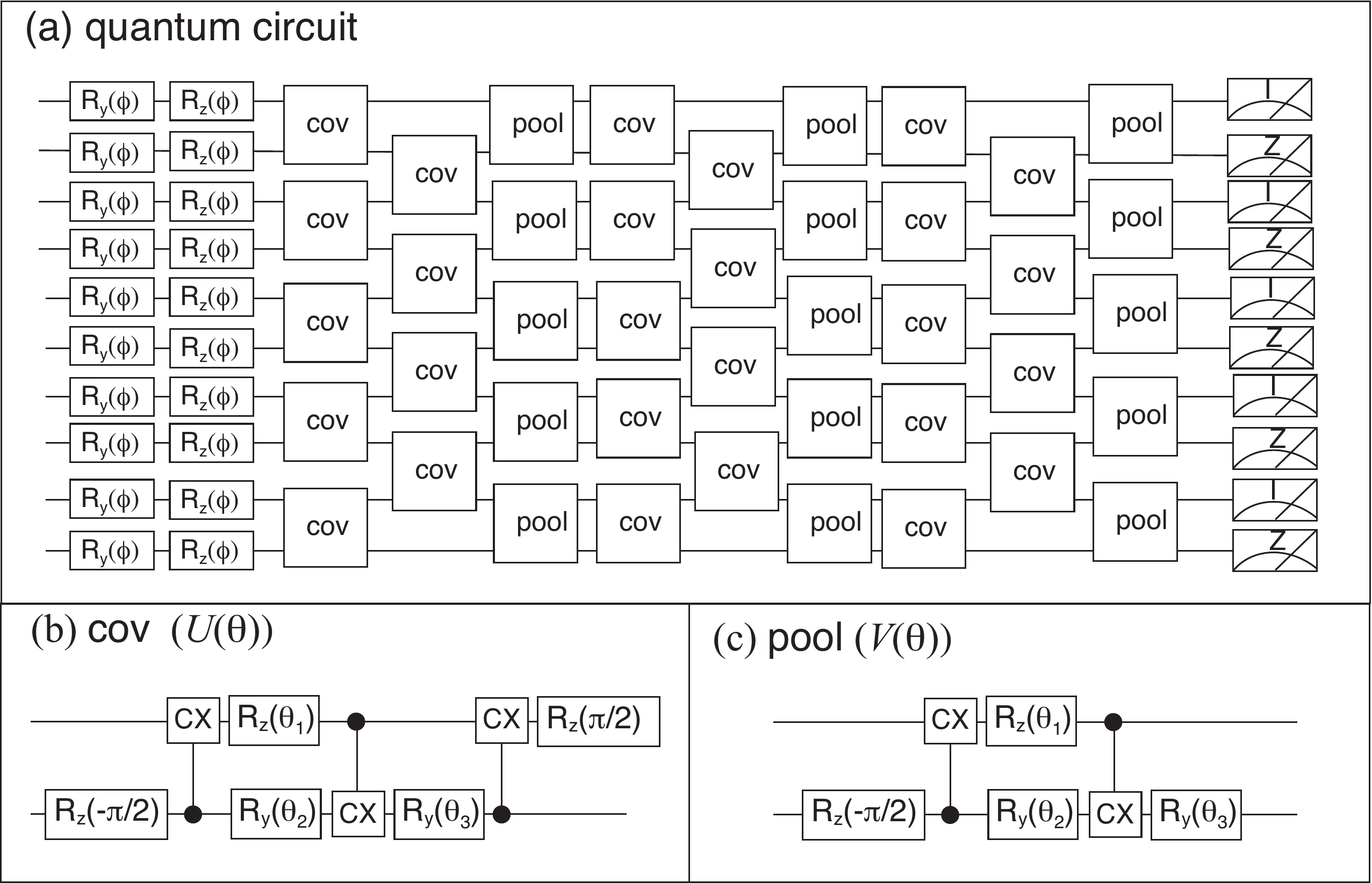}
\caption{ {Quantum circuit.} (a)  A quantum circuit used in Fig. 2. (b) The circuit for the convolutional (cov) layer. (c) The circuit for the pooling (pool) layer.
}
\label{Fig:fig7}
\end{figure}

Figure~\ref{Fig:fig7} plots the 10-qubit quantum circuit used in Fig.~\ref{Fig:fig3}. This circuit includes a rotation layer with $R_y$ and $R_z$ operators, three convolutional layers, and three pooling layers. The quantum circuits for the convolutional $U({\bf \theta})$ and pooling $V({\bf \theta})$ operations are shown in Figs.~\ref{Fig:fig7}(b) and~\ref{Fig:fig7}(c), respectively. Variables are shared in each convolutional and pooling layers. These two operators are also used in Fig.~\ref{Fig:fig10}.

\section{Training results on the FMNIST dataset}\label{sec:cnnloss}
\begin{figure}[h]
\centering\includegraphics[width=0.9\columnwidth]{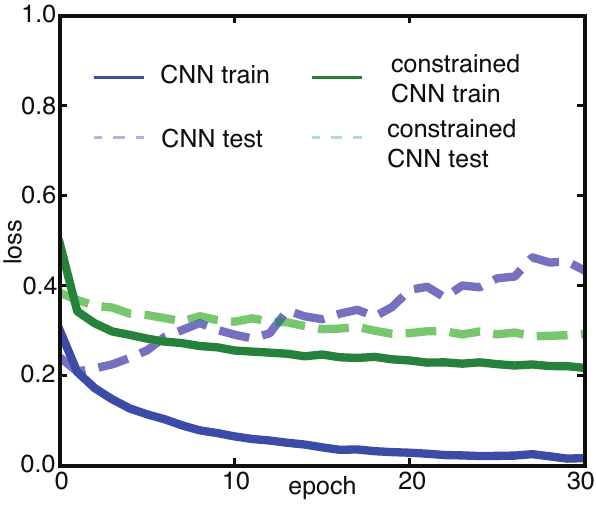}
\caption{Comparison of the categorical cross-entropy loss functions for the CNN and weight-constrained CNN on the FMNIST dataset.}
\label{Fig:fig8}
\end{figure}

Here, we compare the categorical cross-entropy loss functions of the CNN and the weight-constrained CNN for the FMNIST dataset, as shown in Fig.~\ref{Fig:fig8}. The blue solid and dashed curves represent the training and test losses of the CNN, respectively, while the green solid and dashed curves correspond to those of the weight-constrained CNN. The parameters for the combination set are $N=18$ and $r=4$. The blue solid curve decreases continuously with increasing epoch number, whereas the blue dashed curve decreases initially and then increases, indicating overfitting in the CNN. In contrast, both the training and test losses of the weight-constrained CNN decrease monotonically, demonstrating that the overfitting issue is effectively mitigated by the reduced expressibility of the weight-constrained network. Furthermore, we note that the minimal test loss of the CNN is smaller than the test loss of the weight-constrained CNN. Adopting the early-stop method, the optimal CNN model has a larer test ACC compared to the weight-constrained CNN, as shown in Fig. ~\ref{Fig:fig3}(a).

\section{Training results on the MNIST dataset}\label{sec:mnist}
\begin{figure}[h]
\centering\includegraphics[width=0.9\columnwidth]{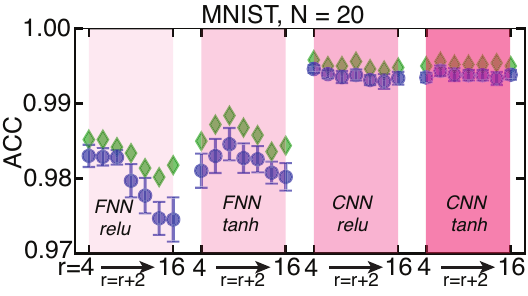}
\caption{Training results on the MNIST test dataset using different weight-constrained neural networks.}
\label{Fig:fig9}
\end{figure}

Figure~\ref{Fig:fig9} compares the training results on the MNIST test dataset for four different weight-constrained networks: FNNs and CNNs employing either ReLU or Tanh activation functions. In the figure, the blue circle denotes the mean ACC, and the green diamond represents the optimal ACC. For the weight-constrained FNN with ReLU, the mean ACC decreases as $r$ increases. In contrast, for the weight-constrained FNN with Tanh, the mean ACC initially increases and then decreases with $r$. However, as these variations fall within the range of static error, we conclude that the mean ACC for the Tanh FNN is effectively independent of $r$. For the weight-constrained CNN, the mean ACC is independent of $r$ for both ReLU and Tanh functions. This is because the preceding layers can adjust the neuron values $x$ in the weight-constrained layer, thereby mitigating the expressivity limitations imposed by the approach.

\section{Different numbers of features}
\begin{figure}[t]
\centering\includegraphics[width=0.95\columnwidth]{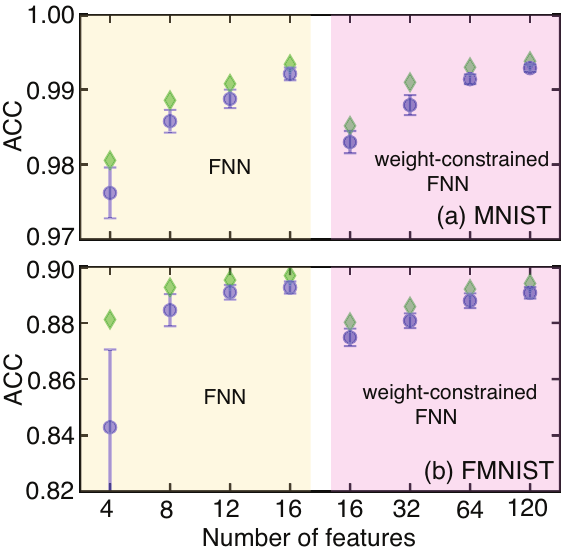}
\caption{The test accuracy (ACC) as a function of the number of features in the first hidden layer for both the standard and weight-constrained fully connected neural networks (FNNs). Results on the MNIST and Fashion-MNIST (FMNIST) datasets are presented in panels (a) and (b), respectively. The mean and optimal results are represented by blue circles and green diamonds, respectively. The weight-constrained matrix is constructed by setting $N=16$ and $r=4$.
}
\label{Fig:fig10}
\end{figure}

Figure~\ref{Fig:fig10} plots the test ACC as a function of the number of features in the first hidden layer for both standard and weight-constrained fully connected neural networks (FNNs). Blue circles and green diamonds represent the mean and optimal results from 16 independent simulations. For the weight-constrained FNN, the weight-constrained matrix is constructed by setting $N=16$ and $r=4$. The results on the MNIST and Fashion MNIST (FMNIST) datasets show that ACC increases as the number of features increases for both networks.  Compared to the standad FNN, the weight-constrained FNN requires more features to achieve a similar ACC. For example, on MNIST, the standard FNN attains a mean ACC of about 0.98 with only 4 features, while the weight-constrained FNN needs 16 features to reach comparable performance. Nevertheless, the weight-constrained FNN uses fewer parameters overall. The FNN with 4 features in the first hidden layer has 3712 weights, while the weight-constrained FNN with 16 features has 1600 weight variables. A similar trend is observed for the FMNIST dataset: the standard FNN with 12 features achieves a mean ACC of about 0.89 using 10496 weights, while the weight-constrained FNN with 64 features reaches a similar mean ACC with 5440 weight variables.

\section{Amplitude encoding}
\begin{figure*}[t]
\centering\includegraphics[width=\textwidth]{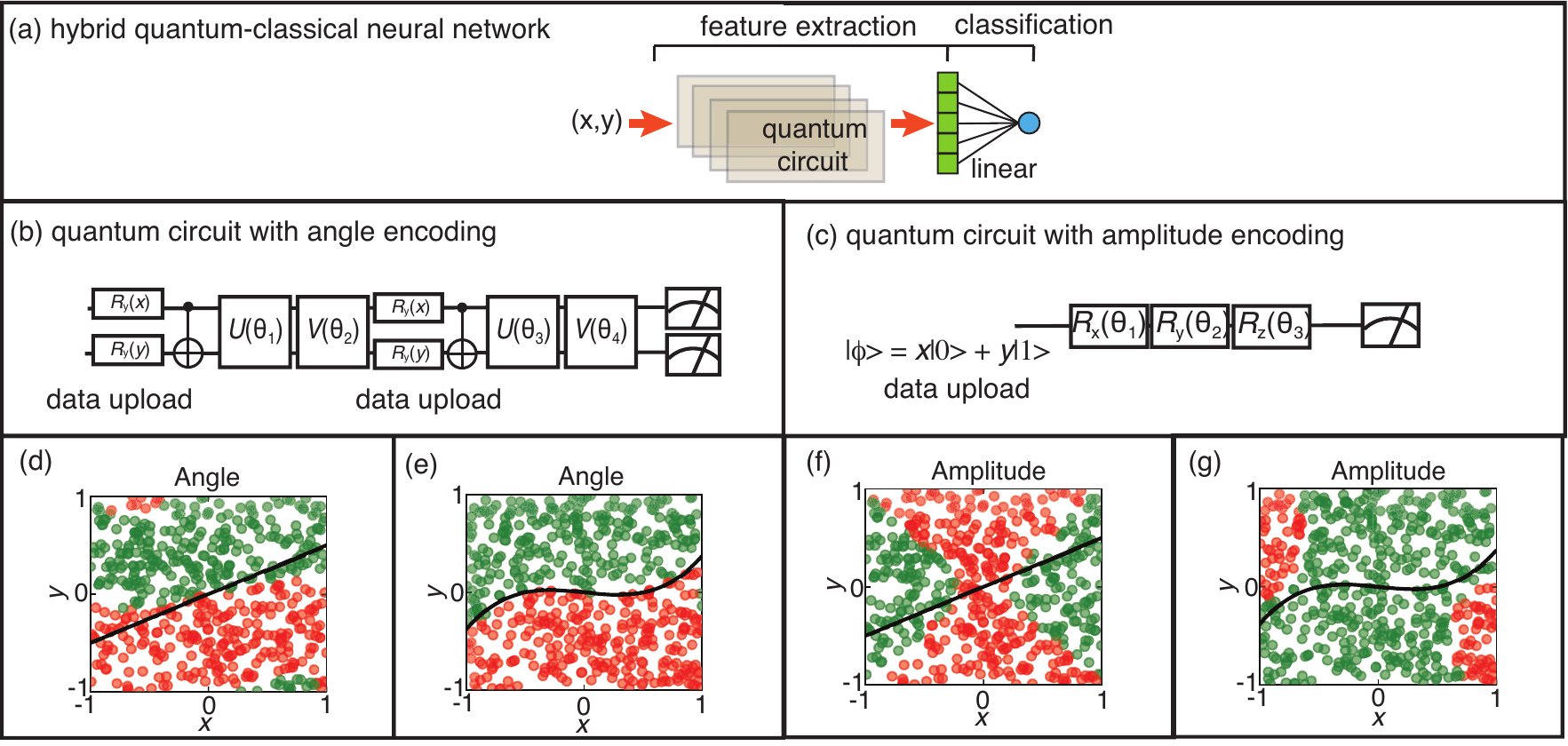}
\caption{Hybrid quantum-classical neural network. 
(a) A hybrid quantum-classical neural network. (b) A two-qubit quantum circuit using angle encoding. The operators $U$ and $V$ are illustrated in Appendix A. (c) A one-qubit quantum circuit using amplitude encoding. (d) Training results of the hybrid network with angle encoding for classifying whether a data point lies above or below the boundary $y=x/2$, shown as a black line. The green color denotes the predicted value above the boundary, while the red indicates the prediction below the boundary. (e) Training results of the hybrid network with angle encoding for classifying whether a data point lies above or below the boundary $y=x(x-0.5)(x+0.5)/2$. (f) Training results of the hybrid network with amplitude encoding, addressing the same task as in panel (d). (g) Training results of the hybrid network with amplitude encoding, addressing the same task as in panel (e).
}
\label{Fig:fig11}
\end{figure*}

The quantum neural network (QNN) with amplitude encoding inherently limits the nonlinear relationship between input and output to a quadratic order, regardless of the complexity of the quantum circuit. This constraint makes it challenging for the QNN with amplitude encoding to extract features that involve three-point or four-point correlations. In contrast, a QNN with angle encoding can more effectively capture these higher-order correlations, as shown in Eq.~(\ref{eq:angle}). To illustrate this limitation, we consider the task of identifying whether a data point is above or below a boundary.

Figure~\ref{Fig:fig11}(a) shows a hybrid quantum-classical neural network (HNN) that we adopt to perform the task, classifying whether a data point is above or below a boundary.
In this setup, the quantum circuit is responsible for feature constructions, while a fully connected classical neural network with a linear activation function performs the classification. The quantum circuits for angle encoding and amplitude encoding are depicted in Figs.~\ref{Fig:fig11}(b) and~\ref{Fig:fig11}(c), respectively. The two-qubit operators $U$ and $V$ are sketched in Appendix A. 
As shown in Figs.~\ref{Fig:fig11}(d)-\ref{Fig:fig11}(g), we classify two different boundaries: $y_1=x/2$ and $y_2=x(x-0.5)(x+0.5)/2$, using angle and amplitude encodings. The black curve denotes the boundary, the green dot denotes the predicted value above the boundary, and the red dot denotes the predicted value below the boundary. The QNN with amplitude encoding struggles to identify both boundaries because the features derived from terms like $x^2$, $xy$, and $y^2$ are insufficient to characterize $y_1$ and $y_2$. However, the QNN with angle encoding successfully identifies the boundary for $y_2$ due to its ability to capture the third-order term (i.e., $x^3$) inherent in the problem. The higher-order term can be obtained from the Taylor expansion of the cosine function. The subpar performance of angle encoding for $y_1$ can be attributed to the excessive nonlinearity imposed by the QNN.

\section{Expressibility}
\begin{figure}[h]
\centering\includegraphics[width=0.99\columnwidth]{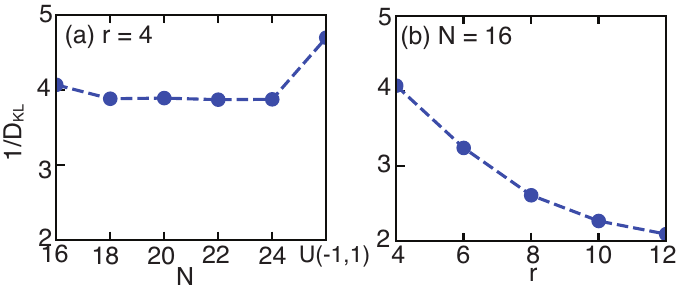}
\caption{(a) The inverse of Kullback-Leibler divergence $D_\text{KL}^{-1}$ of the distribution $p(z)$ as a function of $N$ at $r=4$, where $N$ denotes the number of variables. (b) $D_\text{KL}^{-1}$ of the distribution $p(z)$ as a function of $r$ at $N=16$. Here, the reference distribution is set as the output distribution of a network with uniformly distributed weights.
}
\label{Fig:fig12}
\end{figure}
In this section, we present the inverse of Kullback-Leibler divergence $D_\text{KL}^{-1}$ with $q(z)=1/(2R)$ in Fig~\ref{Fig:fig12}. Here, $R$ is set as 100, which is large enough to ensure $p(R)\approx0$. The result shows that $1/D_\text{KL}$ is nearly independent of $N$ but decreases rapidly as $r$ increases. Furthermore, we observe that $1/D_\text{KL}$ of uniformly distributed weights is larger than that of the weight-constrained network, indicating that the standard network exhibits higher expressibility than the weight-constrained network.

\section{Dataset}
In our study, we adopted four different datasets: MNIST, Fashion MNIST, CIFAR, and a real-world traffic sign dataset derived from the LISA Traffic Sign Dataset~\cite{Mogelmose2012vision} and the Mapillary Traffic Sign Dataset~\cite{ertler2020mapillary}. For each dataset, only five classes were selected as samples.
For the MNIST, FMNIST, and CIFAR datasets, we use 30,000 samples for training and 5,000 samples for testing. The traffic sign dataset has fewer samples, so we use 200 samples for training and 50 samples for testing.

% Bibliography
\bibliography{main}

\end{document}